%% file: 00_main.tex
\documentclass[sigconf, nonacm]{acmart}

\setcopyright{none}             
\renewcommand\footnotetextcopyrightpermission[1]{} 
\AtBeginDocument{%
  }

\def\authnote{0} 
\def\revview{0} 
\input{template/defs}

\usepackage{booktabs}
\usepackage{multirow}
\usepackage[table,xcdraw]{xcolor}
\usepackage{enumitem}

\begin{document}

\title[The Relational Capacity Framework]{The Relational Capacity Framework:\\A Lens for AI-Mediated Relationality}

\author{Emily Tseng}
\affiliation{%
  \institution{University of Washington}
  \country{USA}
}
\author{Patrick Gage Kelley}
\affiliation{%
  \institution{Google}
  \country{USA}
}
\author{Renee Shelby}
\affiliation{%
  \institution{Google Research}
  \country{USA}
}
\author{Ashley M. Walker}
\affiliation{%
  \institution{Google}
  \country{USA}
}

\renewcommand{\shortauthors}{Tseng et al.}

\input{sections/_abstract}




\maketitle

\input{sections/01_introduction}
\input{sections/02_relwork}
\input{sections/03_methods}

\input{sections/04_findings}
\input{sections/05_states}
\input{sections/_rcf}
\input{sections/06_discussion}
\input{sections/_conclusion}

\input{sections/_acks}

\bibliographystyle{template/ACM-Reference-Format}
\bibliography{99_bibliography}

\input{sections/_appendix}

\end{document}

%% file: template/defs.tex
\newcommand{\todo}[1]{\ifnum\authnote=1{\textcolor{red}{[TODO: #1]}}\fi}

\newcounter{mynote}[section]

\newcommand{\thenote}{\thesection.\arabic{mynote}}

\newcommand{\emnote}[1]{\ifnum\authnote=1\refstepcounter{mynote}{\textbf{\textcolor{purple}{$\ll$ET~\thenote: {\sf #1}$\gg$}}}\fi}
\newcommand{\awnote}[1]{\ifnum\authnote=1\refstepcounter{mynote}{\textbf{\textcolor{orange}{$\ll$AMW~\thenote: {\sf #1}$\gg$}}}\fi}
\newcommand{\renote}[1]{\ifnum\authnote=1\refstepcounter{mynote}{\textbf{\textcolor{green}{$\ll$RS~\thenote: {\sf #1}$\gg$}}}\fi}
\newcommand{\pgknote}[1]{\ifnum\authnote=1\refstepcounter{mynote}{\textbf{\textcolor{blue}{$\ll$PGK~\thenote: {\sf #1}$\gg$}}}\fi}

\newcommand{\revadd}[1]{\ifnum\revview=1{\color{blue}{#1}}\else{#1}\fi}

\newcommand{\revout}[1]{\ifnum\revview=1{\color{red}\sout{#1}}\else{}\fi}

\newcommand{\secref}[1]{\S\ref{#1}}

  {\list{}{\leftmargin=0.1in\rightmargin=0.1in}\item[]}%
  {\endlist}

%% file: sections/_abstract.tex
\begin{abstract}
  A growing literature examines the impact of engaging with AI chatbots on our relationality: our ability to form relationships with others, maintain them through life's changes, and exit when needed. Currently, substantial work focuses on purpose-built AI companions (e.g. Replika, Character.AI), with less work investigating socioemotional uses of the more widely-used general-purpose chatbots. In this paper, we present a 100-person investigation into how people engage general-purpose AI for socioemotional needs. Participants' AI uses varied, from practicing specific social skills to seeking additional bandwidth and information as they navigated their social contexts. To explain this heterogeneity, we provide 8 states that describe how AI expands or contracts relationality for people with different levels of skills, bandwidth, and information. We then present a unified conceptual model for the role of technology in mediating human relationality, the Relational Capacity Framework, and outline future research designing appropriate roles for AI in human connection.
\end{abstract}

%% file: sections/01_introduction.tex
\section{Introduction}
\label{sec:intro}

Relationality---our individual and collective ability to create, maintain, and navigate healthy relationships with each other---is foundational to human wellbeing \cite{hart2024exploring}. 
Understanding how emerging technologies shape this capacity is therefore necessary to build systems that limit risks to the social fabric, and support human connection \cite{Zhang2025}. 
Today, as people increasingly turn to general-purpose AI to meet various social, psychological, and emotional needs, such as relationship advice \cite{Tseng2026}, introspection \cite{InnerPond}, and companionship \cite{Zhang2025, Hollis2026}, we face urgent questions about how these systems shape our relationships. 
As with prior technological waves, like email and social media \cite{baym2015personal}, the integration of AI into social life has sparked as much excitement for its connective potential as anxiety over its interpersonal costs. 
Striking the right balance is critical, yet we lack a clear understanding of the mechanisms driving this integration.


Clarity on these mechanisms requires investigating how people actually use AI to manage their relationality. The emerging literature highlights substantial heterogeneity in user practices: people employ AI to edit the tone of text messages \cite{fu2025should}, plan dates, and interpret others’ motivations \cite{Tseng2026}, among others. Furthermore, human relationality is highly contextual. A college freshman building a new network has vastly different relational needs than a new parent experiencing situational time and attention constraints. To ensure AI scaffolds rather than substitutes for human connection, we must understand the precise functions people seek from these systems and how these uses expand or constrain their pursuit of relationality. In this work, we take steps toward understanding this by asking:

\begin{enumerate}
    \item What functions do people seek from AI as they manage their relationality?
    \item How can we explain the heterogeneity we see in how people leverage AI in managing their relationality?
\end{enumerate}
We pursued these research questions through a 100-person mixed-methods study of people who had used general-purpose AI for social and emotional support within the past 30 days. All 100 participants answered a qualitative survey characterizing their use of AI for socioemotional needs, and recorded a think-aloud interaction with their chosen AI system around a typical socioemotional need. We followed up with 23 in-depth interviews to probe why participants bring these needs to AI, and how the technology fits into their broader relationships. From this data, we conducted an abductive analysis integrating our empirical findings with established HCI, psychology, and sociology frameworks. 

We found people are actively managing their relational lives with AI. Many turn to AI to support a broad range of \textit{social skills}, including deciding to initiate or end relationships, regulating emotions, and adjusting their communication to accommodate others or self-advocate. However, social skill support only partially explains this usage. People also use AI systems to manage constrained \textit{relational bandwidth} (encompassing both internal reserves, like  time, energy, or emotional load, and external resources, like the availability of one's social support network), and to acquire the \textit{contextual information} necessary to apply their skills effectively in complex environments. We synthesize these findings into the \textbf{Relational Capacity Framework} (RCF). The RCF conceptualizes relational management as a metacognitive process of monitoring and rebalancing one’s skills, bandwidth, and information, with AI serving as a flexible resource to subsidize whichever component is lacking at the time. To demonstrate the RCF’s explanatory power, we provide a series of case studies mapped across states of relational capacity, illustrating how this variance explains the breadth of AI use for socioemotional needs. 

The RCF provides conceptual footing for designers and researchers to study how people use technology to relate to one another across the life course, offering a lens on tech-mediated relationality beyond the binary of ``connection vs. isolation.'' We close with design recommendations and an agenda for future HCI research aimed at scaffolding human relational capacity. In sum, we contribute:

\begin{enumerate}[label={\textbf{C\arabic*.}}, ref={C\arabic*}]
    \item An empirical account of how people interact with general-purpose AI chatbots to manage relationality, revealing that usage spans social skills support, bandwidth management, and information seeking (\secref{sec:findings}). 
    \item A typology of user states (\secref{sec:states}) demonstrating how variance in people's circumstantial skills, bandwidth, and information explain the heterogeneity of social and emotional AI use.
    \item The Relational Capacity Framework (RCF), a conceptual model that frames tech-mediated relationality as the metacognitive management of skill, bandwidth, and information, mapping the potential risks or benefits of AI use to one’s relationality (\secref{sec:rcf}).
    \item Design recommendations and a research agenda for developing AI systems that scaffold rather than substitute human relational capacity (\secref{sec:discussion}).
\end{enumerate}

%% file: sections/02_relwork.tex
\section{Related Work}
\label{sec:relwork}

Every wave of computing has introduced new forms of connection---and, correspondingly, new anxieties about how technological affordances like instant communication and at-scale connection alter our social fabric \cite{baym2015personal}. At the heart of these anxieties is a concern for \textbf{relational capacity} \cite{Zhang2025}: our individual and collective ability to be in relationship with other people, to create new relationships, maintain existing ones through life’s turns, and exit or downshift when necessary. Relationality is recognized across fields as a pillar of human well-being. \textit{Self-determination} theory positions “relatedness” as one of three innate needs, alongside autonomy and competence \cite{deci2012self}. The \textit{social support} literature shows even the perception of having people to reach out to can physically buffer stress \cite{shumaker1984toward}. On a macro level, interdependence is a necessary element of community resilience \cite{cross2000relational}, particularly during collective crises \cite{lejano2023caring}; and longitudinal evidence from the 85-year Harvard Study of Adult Development identifies relationship quality as the strongest predictor of lifelong health and happiness \cite{hart2024exploring}. In short, how well we relate to others shapes how well we live. 

HCI researchers have long examined computing's impact on relationships across a range of contexts. Subfields focusing on social computing and digital intimacy explore how technologies scaffold interpersonal closeness across distance and time \cite{geeng2025intimacy}, while literature on tech-facilitated abuse highlights how these same systems can weaponize connectivity, granting abusers intimate access to stalk, surveil, and harass their victims \cite{tseng2022careinfrastructures}. 
Between these extremes, HCI has studied how technology mediates relationality in contexts where it is under particular strain: families navigating distance and transition \cite{gupta2026, le2024children}, sexual and romantic partnerships managing communication and conflict \cite{Tseng2026}, caregivers burdened by emotional, temporal, and professional demands \cite{smirti2024caregiving,liu2025regulation, ming2023go, solano2026sharing}, and workplaces where professional and personal relational demands collide~\cite{akridge2026oh}. In a more philosophical register, relationality is constitutive of the care ethics \cite{lejano2023caring} that underpin HCI scholarship on societal resilience and mutuality \cite{semaan2019routine, petterson2025expanding}. Taken together, this diverse body of work reveals a field deeply concerned with preserving and improving our ability to connect with each other.

LLM-enabled conversational systems represent the latest chapter in this lineage, yet they also introduce new affordances of relationality. Unlike prior communication technologies that primarily mediated connection between people, conversational AI can act as relational interlocutor itself. As these technologies are uniquely responsive, personalized, and on-demand \cite{chandra2025lived}, they actively shape what users bring to their human relationships \cite{ho2018psychological}. This socioemotional use is pursued across a growing range of products. Companion AI systems, such as Character.AI \cite{characterai} and Replika \cite{replika}, are designed to foster emotional closeness through configurable personas \cite{lee2026large} while therapy-oriented apps like Ash operate in accordance with psychotherapeutic principles \cite{ash}. However, growing evidence suggests that these social and emotional interactions extend beyond purpose-built apps, where users  appropriate general-purpose AI systems designed primarily for information seeking and productivity for relationship advice, introspection, emotional processing, and social support \cite{claudesocialsupport}. 

As historical perspective portends \cite{baym2015personal}, this latest wave of tech-mediated connection has brought renewed anxieties over the future of human relationality \cite{zhang2026interaction,Turkle2024Who}. A dominant concern is the “substitution hypothesis,” the fear that AI interactions will replace human connections \cite{zhang2026interaction}. Psychologist Sherry Turkle warns of an emerging artificial intimacy, suggesting that AI could ultimately make us lonelier by conditioning us to expect frictionless relationships \cite{Turkle2024Who}, thereby atrophying the social skills required for messy human relationships \cite{wang2026futuring}. Furthermore, researchers have documented explicit harms, demonstrating how adversarial or poorly designed companion systems can actively degrade users’ relational capacities \cite{Zhang2025,ibrahim2026sycophantic}. 
As AI’s long-term impact on our social fabric unfolds, we urgently need research identifying the specific mechanisms by which it influences relationality to design better futures.

Despite sustained interest in technology and relational life, existing frameworks for understanding what drives relational capacity remain largely siloed within psychology, and primarily focus on specific intra- and interpersonal skills \cite{buhrmester1988five,carver1989assessing, schoon2021towards, bandura1982self}.  The Interpersonal Competence Questionnaire, for instance, measures relational ability as discrete, learnable competencies such as initiation, self-disclosure, self-assertion, emotional support, conflict management \cite{buhrmester1988five}. Carver’s inventory of coping skills evaluates how people adaptively and maladaptively respond to stress \cite{carver1989assessing}, while DOMASEC decomposes relational functioning into intrapersonal, interpersonal, and task-based skills \cite{schoon2021towards}. These approaches suggest relational capacity is a matter of skill: what Bandura’s framework of self-efficacy frames we can improve through deliberate practice and encouragement \cite{bandura1982self}. 

However, relational capacity involves more than what one can practice and control; it is fundamentally dyadic and contextual \cite{cross2000relational}. The interpersonal process model of intimacy shows that felt closeness depends on one's own disclosures and on a partner's reciprocal disclosure and perceived responsiveness \cite{reis2018intimacy}. Uncertainty reduction theory establishes that navigating relationships requires active information seeking to understand shifting expectations, cultural norms, and situational dynamics \cite{kramer1999motivation}. In HCI, this resonates with sensemaking research \cite{pirolli2011introduction}, which shows effective action in complex environments depends on the ability to gather, interpret, and organize contextual information \cite{russell1993sensemaking}.

Even when people possess both the necessary skills and contextual information, their relational capacity can be constrained by finite resources. In the literature on human well-being broadly, the capabilities approach advanced by Sen \& Nussbaum posits that a person may possess the capability to do or be what they wish (e.g., everyone can become educated), but require resources to convert those capabilities into functionings (e.g., everyone needs access to quality education to be well-educated) \cite{nussbaum1993quality}. The care literature offers hints to what those resources might look like in the context of relationality: Hochschild’s foundational work on emotional labor \cite{hochschildmanaged} and the caregiving literature on burden and fatigue \cite{marie2017family, ren2024emotional} make clear that relationality depends on depletable resources.

What is missing then, is a framework that unifies these mechanisms: the skills that enable relational action, the information that contextualizes it, and the bandwidth that sustains it. Moreover, we need a lens to understand how technology might mediate each of these dimensions. Clarifying these effectors of relationality is essential for designing technologies  that actively scaffold positive human connection throughout the lifespan. Such a vision extends HCI’s longstanding interest in human-machine complementarity \cite{zerilli2019algorithmic} in cognitive work into the messiest and most joyful parts of the human experience: how we relate to each other. We place our work on this frontier.

%% file: sections/03_methods.tex
\section{Methods}
\label{sec:methods}

To understand how people use general-purpose AI to navigate and manage their interpersonal lives, we conducted a qualitative mixed-methods study. The study comprised an unmoderated survey, an embedded think-aloud task centered on a real-world relational need, and semi-structured follow-up interviews. These methods enabled us to collect open-ended insights without researcher influence \cite{braun2021online}, and expose participants’ underlying cognitive processes when interacting with an AI agent \cite{ericsson2017}. To complement the survey and think-aloud task, we conducted 23 follow-up interviews to elicit more in-depth context around how participants made decisions about using AI for social and emotional support, as well as how their usage of AI for social and emotional support has evolved \cite{brinkmann2013qualitative}.

\subsection{Study Design}
In this work, we investigate how participants characterized AI’s role in managing their relationality. We specifically focused on study components that asked participants to detail:

\begin{enumerate}
    \item Historical and current practices of using AI for relational needs, and how those experiences intersect with how they broadly manage relationships with other people.
    \item Reflections on the specific functions AI serves in these moments (e.g., how and why they turn to AI for a relational need), and the lessons they have learned regarding AI-mediated relationship management. 
    \item How AI is situated alongside other ways participants get support.
\end{enumerate}

During the survey and think-aloud interaction, participants contributed written responses, as well as audio and video recordings documenting real-life instances where they turned to AI to navigate an interpersonal hurdle. The subsequent in-depth interviews expanded upon these themes, allowing researchers to probe heavily into why participants selected AI over or alongside other support systems, and how these practices shaped their broader relational ecosystem. 

\subsection{Participant Recruitment}

We recruited 100 US-based adults via the dScout research platform, inviting a subset of 23 for one-hour semi-structured interviews. To be eligible, participants had to be 19 or older, reside in the U.S., speak English, and have used a general-purpose AI chatbot (e.g., ChatGPT, Gemini, Claude) for emotionally salient or interpersonal conversations within the prior 30 days. Data collection began in February 2026 with follow-up interviews concluding in March 2026. 

During screening, prospective participants reported their specific AI use cases. Eligible participants indicated using AI for at least one of the following: interpersonal advice or coaching, self-discovery and reflection, managing your emotions, specific coping skills, role playing offline interactions, AI friendship or companionship. We also gathered data on their frequency of use, tenure with the technology, and willingness to discuss sensitive topics that commonly arise in social and emotional contexts (e.g., challenging work situations, interpersonal conflict, or regulating complex emotions such as stress, anxiety and loneliness).  

From a pool of 527 qualified respondents, we purposively sampled 100 participants to ensure a balanced distribution across usage frequencies and diverse demographic backgrounds. See Table \ref{tab:demographics} for details on participants’ self-reported demographics. Participants provided informed consent and were reminded that they were free to withdraw at any time. Upon completion of the survey, they were compensated USD\$30. 

\input{tables/demographics}

Before exiting the survey, participants were asked if they were willing to be contacted again for a potential follow-up interview. From this pool of possible participants, we invited a sample that represented a range of experiences across high, medium, and low frequency chatbot users. Eligible follow-up interview participants were contacted through the dScout platform and invited to complete a second consent form, for a 1-hour interview for which they would receive an additional USD\$75 compensation. All interviews were conducted through the dScout platform by three of the authors and audio-recorded to ensure transcription accuracy. All participants were informed that they could terminate the interview, take a pause, and skip any questions, and still receive the full compensation.

\subsection{Research Ethics and Positionality}

In studying relationality and the use of AI around sensitive personal topics, we were conscious to create space in our research methods for participants to share about what can be intimate and/or stigmatized practices. As a research team, we have each had over 10 years of experience with at-risk users, including training and practice in trauma-informed computing, and research with queer communities and survivors of domestic and gender-based violence. These experiences helped us structure our research with an expansive approach to potential experiences of relationality. Additionally, while our study did not center crisis or mental health, we understood there might be occasional overlap between relationality and these topics. Out of an abundance of caution and taking a trauma-informed approach, we provided participants with a list of mental health resources upon completion of the survey, including 988 Suicide \& Crisis Lifeline, the National Alliance on Mental Illness (NAMI), the Substance Abuse and Mental Health Services Administration (SAMHSA), and Crisis Text Line. 
We additionally paraphrased and pseudonymized all participant quotes throughout this manuscript.
Our study protocol was approved by our Institutional Review Board in January 2026 (Protocol identifier redacted for anonymity).

\subsection{Data Collection and Analysis}
For this paper, the corpus for analysis included the complete set of survey responses (100 responses, 41 questions), the think-aloud interaction, and interview transcripts (machine-transcribed and manually corrected for accuracy). Recognizing that self-reported practices are susceptible to social desirability bias and limited self-knowledge, participants may overestimate their skills or underreport information deficits. We partially mitigated these limitations by triangulating across survey responses, think-aloud interactions, and interview probes. 

We conducted an abductive analysis \cite{Timmermans2012, Tavory2014abductive} to simultaneously make sense of participants’ self-reports and cohere different theoretical lenses (summarized in S2). First, all authors familiarized themselves with the corpus, noting patterns and recurring themes based on concepts drawn from the larger literature on relationality. Initial team discussions established a starting theoretical lens focusing on how participants manage their social skills. To test how well this lens could explain the data, two authors conducted focused coding on 25 survey responses. In line with abductive analysis, we then examined the  findings that social skill management could not explain. Through iterative discussions, the broader team identified two additional lenses required to explain the data: resource management (bandwidth) and information-seeking (information). Two authors then applied this triad of lenses — skills, bandwidth, and information — to code the entire corpus of 100 survey responses and 23 interview transcripts. Finally, through multiple rounds of synchronous and asynchronous team discussions, we developed the themes that structure Sections 4, 5, and 6.

%% file: tables/demographics.tex
\begin{table}[t]
\centering
\caption{Participant Demographics (N = 100)}
\label{tab:demographics}
\begin{minipage}[t]{0.48\textwidth}
    \centering
    \begin{tabular}[t]{lr}
        \toprule
        \textbf{Category} & \textbf{n} \\
        \midrule
        \textbf{Gender} & \\
        \quad Woman & 49 \\
        \quad Man & 47 \\
        \quad Non-binary & 4 \\
        \addlinespace
        \textbf{Age Group} & \\
        \quad 18--24 & 19 \\
        \quad 25--34 & 24 \\
        \quad 35--44 & 26 \\
        \quad 45--54 & 20 \\
        \quad 55--64 & 9 \\
        \quad 65--74 & 2 \\
        \addlinespace
        \textbf{Duration of Socioemotional AI Use} & \\
        \quad Less than three months & 27 \\
        \quad 3-6 months  & 32 \\
        \quad 7-12 months  & 24 \\
        \quad More than a year  & 17 \\
        \bottomrule
    \end{tabular}
\end{minipage}\hfill
\begin{minipage}[t]{0.48\textwidth}
    \centering
    \begin{tabular}[t]{lr}
        \toprule
        \textbf{Category} & \textbf{n} \\
        \midrule
        \textbf{Race/Ethnicity} & \\
        \quad White & 40 \\
        \quad Black or African American & 22 \\
        \quad Multiracial / Two or More Races & 14 \\
        \quad Hispanic or Latinx & 10 \\
        \quad Asian & 9 \\
        \quad Other & 2 \\
        \quad Prefer not to say & 3 \\
        \addlinespace
        \textbf{Geographic Area (US Region)} & \\
        \quad South & 44 \\
        \quad Midwest & 21 \\
        \quad Northeast & 19 \\
        \quad West & 16 \\
        \bottomrule
    \end{tabular}
\end{minipage}
\end{table}

%% file: sections/04_findings.tex
\section{Findings 1: Functions for AI in Relationality}
\label{sec:findings}

Our analysis identified participants use AI to manage their relational lives along three dimensions: social skills, relational bandwidth, and contextual information. These dimensions operate as dynamic and contingent resources: from circumstance to circumstance, people may find themselves highly resourced in one area but temporarily depleted in another. We first detail skill-oriented uses of AI (\secref{sec:findings-skills}), then show how fluctuating bandwidth and information explain what skills alone cannot (\secref{sec:findings-bandwidth-information}).

\subsection{AI helps people deploy social skills}
\label{sec:findings-skills}

\input{tables/skills}

Participants turn to AI seeking support with common social skills that map onto established frameworks of interpersonal and intrapersonal competence \cite{buhrmester1988five, carver1989assessing, nussbaum1993quality}. Across our data, we identified eight skills spanning both domains (see Table \ref{tab:skills}). AI functioned to support \textit{mastery} \cite{bandura1982self} by teaching undeveloped skills, providing low-stakes environments for rehearsing skills, or encouraging confident execution. Importantly, we do not claim AI successfully facilitated mastery; we claim only that participants sought such support. 

\subsubsection{Interpersonal Skills} Participants sought AI’s help calibrating the tone or tactics of their \textbf{communications} to either accommodate others or self-advocate. To accommodate, participants wanted to align their words with social norms (e.g., sounding more ``professional'' or ``mature''), or to bridge the gap between their intentions and how those intentions might land. Several participants described situations, often professional, in which they wanted to ensure their communication was kinder or less stressful for the recipient: Shannon asked AI to soften a message, while Tomas valued learning why his phrasing ``\textit{might be misconstrued.}'' Caleb used AI ``\textit{to better formulate my apologies, and protect my partner from me saying something I don’t mean}.'' He called it ``\textit{revolutionary advice}'' because it taught him new communication skills around apology and repair, rather than simply resolving an immediate conflict. Conversely, participants who lacked the language or confidence to assert their needs used AI to self-advocate. Quinn asked AI how to confront lab partners who were not pulling their weight. Samira, a woman in the military, often had trouble relating to her coworkers' wives, who would sometimes assume---in error---that she was interested in their husbands. She asked her AI chatbot, ``\textit{how do I reach out to these people and make it clear that I’m not interested in their husbands, I want to be their friends as human beings, even though our careers are different?}'' Reflecting on this use of AI, Samira noted it felt ``\textit{sexist}'' to reach out to AI with this problem, but it was able to supply answers, teaching her ``\textit{how to approach scary women, essentially}.'' Whether accommodating or self-advocating, participants knew their goals but relied on AI to produce the necessary language.

\textbf{Interpretation} (i.e., understanding others’ thoughts, actions, and emotions) precedes effective communication. When navigating disagreements, participants turned to AI for external frameworks to make sense of confusing behaviors. Grace struggled with her boyfriend’s erratic sleep schedule caused by taking his ADHD medication late in the day. To understand his perspective, she asked AI: ``\textit{What is it like to have ADHD? It's like he has the motivation, but he just doesn’t have the gas to do it.}'' AI directed her to a Reddit community for partners of people with ADHD, which she found helpful. Zachary faced a different interpretive gap regarding how his own behavior landed. After his wife claimed he ``\textit{never listens},'' he asked AI to help him figure out why. The tool reframed his habits, showing him that ``\textit{when I thought that I was listening, I actually wasn’t. I was preparing for my rebuttal.}'' In both cases, AI surfaced perspectives participants could then practice adopting in their real relationships.

Where interpretation decodes another person's behavior, \textbf{judgment} determines the response by gauging what is normal, what is serious, and what warrants a reaction. Participants who already understood the facts of their situation turned to AI to supply evaluative criteria regarding whether to initiate, intensify, downshift, or even end a relationship. Caleb asked, ``\textit{how do I know when it’s a good idea to take a break in a relationship?}'' Fernando sought to assess severity: ``\textit{how serious could I consider a friction between my partner and I?}'' Weighing an irreversible decision, Adam asked whether to ``\textit{complete the divorce or should I do something else}.'' These queries cast AI in an advisory function, treating the system as a repository of relational best practices, social norms, and universal ``\textit{truths}'' to encourage confident decision-making. 

While communication skills focus on what to say, \textbf{vulnerability} (i.e., asking for help, trusting others, and letting oneself be known) centers on whether to speak at all. Some participants lacked confidence in gauging when, how much, and to whom to disclose. Camila avoided burdening her husband because he was ``\textit{always stressed from work},'' fearing her problems would seem trivial. Bilal similarly withheld his feelings from his close ties ``\textit{for fear of being judged}.'' Others hesitated to reach out to their social supports late at night, or when they doubted they would receive adequate support. These participants did not lack the desire to be vulnerable; rather, the barrier was the perceived cost of doing so  \cite{buhrmester1988five}. 

Finally, participants turned to AI to effectively \textbf{provide support} to people they care about---what Buhrmester et al \cite{buhrmester1988five} describe as the interpersonal competence of ``providing emotional support and advice.'' Several participants desired to be supportive but lacked the skill. Adam explained that the stress of being a physician caused conflict with his wife, as ``\textit{all the pressure of the children is on her.}'' To Adam, AI advice was successful in helping him overcome the limitations imposed by his job: ``\textit{I believe that it’s helping me a lot with how to care more about her, how to surprise her, how to reduce our arguing, and how to support her even if I'm not home at all.}'' Rochelle described a similar orientation: ``\textit{I don’t really reach out to AI for emotional support for me, but more so guidance... What I should say in this situation to a friend who’s dealing with this, that, or the other?}'' The skill gap was knowing what to do with the relational care they already felt, with AI functioning to teach concrete supportive actions.  

\subsubsection{Intrapersonal Skills}
Alongside efforts to bolster interpersonal skills, participants described a range of intrapersonal skills to manage themselves. \textbf{Self-soothing} is foundational, as relational situations regularly provoke strong emotional responses that, unmanaged, can escalate conflict or foreclose engagement entirely \cite{leen2004behavioral}. Amidst arguments, bad news, or rejection, participants reached for AI as a first response to manage their physiological state before re-engaging. Finley, furious after a close friend stood him up for the fourth time, explained: ``\textit{I obviously did not want to yell at my friend for a relatively minor offense. So, I talked to AI to help calm myself down, to either distract myself or to move through my anger, and to think of it rationally.}” AI suggested a non-confrontational conversation, prompting ``\textit{a genuine shift in how I’ve processed interactions with others}.'' Lakesha’s use was more targeted: ``\textit{I wanted help practicing a specific grounding technique... I felt empowered.}'' In both cases, participants reached for AI to work through emotions before re-engaging. 

Outside of acute emotional moments, participants sought help with self-care: understanding what behaviors best support one’s body and mind, and taking actions to sustain them. These needs arose when relational strain compromised participants’ ability to function. Mary, navigating a crisis involving her daughter and child protective services, had stopped sleeping: ``\textit{I’ve been really depressed lately and exhausted… what types of things can I do to make myself feel better, because I can’t sleep at night?}'' Her chatbot responded with specific techniques that she planned to continue exploring. Others asked for AI’s support making to-do lists, planning commutes and vacations, or handling chronic health conditions and new diagnoses. In these cases, participants sought self-care to sustain their capacity to show up for the people in their lives. 

Lastly, participants sought AI’s help with \textbf{self-reflection}: understanding one’s own thoughts, actions, and emotions well enough to act on them relationally. Finley described AI as ``\textit{the first entity to suggest that a lifetime of chronic malnutrition, bullying, assault, and poverty might have something to do with many of the problems I was experiencing in my life.}''  Reflecting on arguments with his girlfriend, Khoa sought to ``\textit{see where was the trigger point, because most of the time I don’t realize what I did was wrong.}'' Others needed fundamental help organizing their thoughts. Vincent described AI helping him to ``\textit{put names to things I’m feeling,}'' noting it ``\textit{loops me back into reality… because I have a tendency to be hard on myself.}'' For these participants, AI’s function was to make their internal states legible enough to act on in their relationships.  

\subsection{Beyond skills, AI helps people manage relational bandwidth and contextual information}
\label{sec:findings-bandwidth-information}
The lens of social skill did not completely explain all AI interactions in our study. While many participants sought to develop specific skills, others used AI in ways that bypassed exercising them all together. In these circumstances, participants were not actively avoiding skill mastery, but managing two additional resource challenges within relationality: limited relational bandwidth and a need for contextual information. 

\subsubsection{Managing bandwidth, in others and oneself.} 
\label{sec:findings-bandwidth}
Bandwidth in relationality operates on two fronts: internal bandwidth (the emotional energy, time, and physical capacity an individual has to engage) and external bandwidth (the structural and social support available to them in a given moment. While relying on AI for relational needs might suggest skill avoidance or eventual atrophy, our data points to a different driver: limited bandwidth. Rather than lacking a skill or actively avoiding its development, participants faced circumstantial resource constraints \cite{hobfoll1990conservation} that temporarily restricted their capacity to execute skills they already possessed.  Exogenous pressures, like juggling parenting, high-intensity work periods, or supporting grieving loved ones, severely constrained participants’ relational bandwidth. For instance, Courtney, a well-networked empty nester sought AI’s help after her son-in-law died in a car accident. She needed help coping with her own grief while figuring out what to say to her daughter. Otherwise socially adept, she was not avoiding communication; she simply needed support balancing severe circumstantial constraints to enact a communication skill she already had.

These bandwidth limitations typically manifested as overlapping constraints on stress, time, and available social support. Time constraints drove participants to AI to quickly craft communications, while stress constraints prompted them to seek self-soothing. Finally, constraints on wider social supports drove participants to seek help with vulnerability. Several reported they had tried leaning on families or friends, only to learn the hard way that their networks were unreliable or unable to offer the necessary support.  

\subsubsection{Managing contextual information.} 
\label{sec:findings-information}
Beyond bandwidth, participants needed information to apply the right skill. Rather than teaching, rehearsal, or encouragement toward a skill, they needed to understand their situation well enough to deploy skills they already had: a need akin to appraisal and informational support~\cite{goldsmith2015normative}.

Even seemingly logistical information needs were tied to managing relationships. When facing complex bureaucracies like the American healthcare system or unfamiliar transactions like selling a used car, participants needed a factual grounding to effectively navigate the people within those systems  (e.g.,  doctors, insurance agents, buyers). For example, Haley, a new landlord, used AI to ensure her tenant communications met professional norms and legal obligations. Here, AI functioned as a synthesized information source, providing the baseline context required to successfully manage a specific relational role. Many described their use of AI in these situations as akin to how they would use Reddit or search engines, simply a window into the best advice available on the Internet.

Other informational needs  were highly personal, stigmatized, or tied to unfamiliar cultural and organizational norms. For instance, participants turned to AI to navigate conversations with child protective services, or for advice on how best to empathize with a friend who was upset about their own socioeconomic prospects. Cameron, seeking to support a family member struggling with alcohol, described the information as “\textit{not surprising, it was just mostly helpful information guiding me on the steps that I should take with dealing with someone who has a drinking problem.}”  In these cases, participants emphasized they knew generally how to handle themselves relationally. They simply faced a specific situation that stymied their skill. AI’s function was to supply the situational knowledge that allowed existing skill to become effective action. 

%% file: tables/skills.tex
\begin{table*}[]
\centering
\begin{tabular}{@{}lll@{}}
\toprule
\textbf{Domain} & \textbf{Skill Type} & \textbf{Description} \\ \midrule
 & \cellcolor[HTML]{EFEFEF}Communication & \cellcolor[HTML]{EFEFEF}Adjusting tone or tactics to accommodate others or  self-advocate. \\
 & Interpretation & Understanding others' thoughts, actions, and emotions \\
 & \cellcolor[HTML]{EFEFEF}Judgment & \cellcolor[HTML]{EFEFEF}Gauging whether to begin, intensify, or downshift a relationship. \\
 & Being vulnerable with others & Recognizing a need for help and reaching out to others. \\
\multirow{-5}{*}{Interpersonal} & \cellcolor[HTML]{EFEFEF}Supporting others & \cellcolor[HTML]{EFEFEF}Recognizing when others need help, and providing it effectively. \\ \midrule
 & Self-soothing & Calming oneself from heightened states of emotion or distress. \\
 & \cellcolor[HTML]{EFEFEF}Self-care & \cellcolor[HTML]{EFEFEF}Taking actions to sustain physical and mental health. \\
\multirow{-3}{*}{Intrapersonal} & Self-reflection & Understanding one’s own thoughts, actions, and emotions. \\ \bottomrule
\end{tabular}
\caption{Domains of Relationality Management and Corresponding Skill Types. Across interpersonal and intrapersonal domains, our data show evidence of AI supporting people’s mastery \cite{bandura1982self} of social skills (\secref{sec:findings-skills}). 
}
\label{tab:skills}
\end{table*}

%% file: sections/05_states.tex
\section{Relationality States}
\label{sec:states}

We describe the three dimensions of how people manage their relationships as skill, bandwidth, and information. Together, these dimensions shape a person’s relational capacity in any given situation. In this section, we map the eight states of capacity sourced from our data, representing the various high and low configurations of these dimensions (summarized in Table~\ref{tab:states}). These states are temporary rather than permanent traits; people shift between them or experience multiple simultaneously based on life stages, daily stressors, and social environments. By examining participants’ situational realities through these configurations, we illustrate how these factors shape what is needed from AI. For each state, we detail how AI appeared to expand or contract their relationality. 

\input{tables/states}

\subsection{RS1: low skill, low bandwidth, low information}
In this state, people face deficits across all dimensions of relational capacity. They lack the structural and socioemotional resources to marshal in their situation, as well as the contextual knowledge and skills to handle it well. 

Diamond's AI use illustrates how compounding deficits can manifest into technological dependency for problem-solving, self-management, and decision-making. Working as a night nurse assistant, her time, energy, and human networks are constrained: ``\textit{I don’t really have that many people that I talk to... I don't have friends or family members I can tell my problems to.}'' Diamond turns to AI to manage daily life tasks, from paying rent and handling her dog's separation anxiety to getting dressed and decorating her apartment. Lacking the skills to act on information provided by the chatbot, she relies on it to execute decisions: ``\textit{That's why I go to AI, to tell them what's going on so they can give me responses.}'' She even resents AI features that require her to exercise choice, noting, ``\textit{I really dislike picking because I don't know which one to pick ... I just would like them to just show me a response ... I don't even know how to control my life.}'' 
What results is a state where Diamond’s AI entirely directs her actions, from what to wear to how to take care of her pet to how to manage her emotions.

Conversely, people in this state can actively use AI to build capacity. Monique suffers similar resource deficits, exacerbated by her reluctance to seek reciprocal human support. Feeling ``\textit{overwhelmed and overstimulated with marital issues},'' she turns to ChatGPT because ``\textit{it never makes me feel judged}.'' Unlike Diamond, Monique uses the technology to cultivate her interpersonal and self-management skills. During ``\textit{very dark days}'' contemplating divorce, AI helped her transition from ``\textit{suffering in silence}'' to developing ``\textit{a reasonable plan to work through my feelings, hold difficult conversations, and a strategized exit plan if [the] communication process didn’t work out.}'' By using the technology to locate pro bono counseling in her area, she illustrates how AI can supply necessary information and skill-building to expand relational capacity. 

In this state of compounded deficits, a person’s relational capacity is vulnerable to stasis. They lack the structural resources (bandwidth) and situational context (information) to actively monitor and regulate their own problem-solving. People in this state thus risk delegating control of their relationality to the AI. If the technology does not actively intervene to teach skills and build capacity – as it did for Monique — the user risks cementing a dependence on the machine.  

\subsection{RS2: high information, low bandwidth, low skill}
In this state, people possess high contextual knowledge about their social environment, but lack the tactical ability and structural or socioemotional resources to navigate the situation. They understand the situation and articulate their relational goals, but are too depleted by environmental stressors to execute their intentions.  

Brittany illustrates how AI can provide tactical support when bandwidth is severely taxed. A stay-at-home mother to three children in North Dakota (including a newborn and a child with a disability), she cannot afford therapy and actively avoids seeking support from her network: ``\textit{I don’t wanna burden my husband after he comes home from a long day with all of my inner thoughts and stuff}'' and ``\textit{my friend runs a daycare so she has her own crap going on.}'' Brittany holds clear relational goals, like self-regulation and healthy conflict management, but lacks the skills to meet them when daily stressors compound: ``\textit{the dog pees on the floor, the baby walks through it...I can feel my tether tightening and it's about to snap.}'' She turns to AI for immediate coping skills, such as breathing exercises and perspective-taking to understand her husband's point of view. By seeking scripts to address issues ``\textit{in a productive way versus just yelling at him},'' Brittany works to align her intentions with her capabilities, communicating where she previously ``\textit{would've just sat there and been sad.}''

Tasha experiences similarly constrained bandwidth, and relies predominantly on self-directed coping mechanisms like journaling and self-talk, emphasizing, ``\textit{I really don't talk to other people that I really want to express deep down and really show my vulnerable side}.'' Working in customer service, she squeezes her emotional processing into brief work breaks to ``\textit{get cheered up...and get some things off my chest.}'' Tasha is well-aware of her network’s limitations, noting that ``\textit{people [are] not going to take it the right way, people ain't going to listen.}'' While she recognizes her need for validation, she struggles with epistemic friction that minimizes her own feelings, noting that without AI ``\textit{you feel like you are being overreactive about things.}'' She relies on AI for cognitive reframing, asking it to ``\textit{see different point[s] of views}'' and provide ``\textit{resources to improve yourself.}'' 

Relational capacity is precarious in this state because a person’s contextual knowledge outpaces their structural and socioemotional resources. While people in this state possess the awareness to articulate their goals, their depleted bandwidth hinders their ability to execute them. Consequently, they turn to AI for immediate skills. The risk here is that if the AI does not scaffold the interpersonal skills required to address the root causes of their stress, it may inadvertently help users tolerate unsustainable structural or relational environments, prioritizing coping over necessary change. 

\subsection{RS3: high bandwidth, low information, low skill}
In this state, people are highly resourced with strong support networks. However, they are encountering unfamiliar life phases, novel situations, or specific interpersonal dilemmas where they lack the contextual knowledge to understand what is happening and the tactical tools to respond appropriately. 

Silvia, a lawyer, operates from a well-resourced state. She possesses immense structural support, including a psychologist of eight years, a physiotherapist, a twin sister, and a large friend group. However, she struggles with setting boundaries with her fiance. She shared, ``\textit{I had a fight with [him] because I feel he doesn't have empathy for me... I work 40 hours and I get home and I'm all stressed and tired. And he starts begging that he wants me to cook food for him ... I don't know what else to do to put limits [on] him.}'' Lacking the tactical skill to enforce a boundary and the contextual information to understand his behavior, she uses AI to generate communication scripts and ``\textit{a meditation exercise... a phrase that I could repeat to myself... a sentence of something, the way I should tell him how I feel.}'' Problematically, the AI reframes his lack of empathy as a mere gendered cognitive limitation shaped by ``men’s'' inability to multitask, leading her to conclude, ``\textit{I think we are all selfish.}'' As Silvia lacks information about his behavior and has stopped discussing him with her interpersonal network, she accepts the AI’s reframing, masking a potentially abusive situation. 

Conversely, Camila, an assistant store manager, describes her AI use as ``\textit{extra support}'' alongside her robust human network, noting, ``\textit{I’m an open book. If anything I probably overshare.}'' Preparing to negotiate a raise, she feels she lacks the contextual information to navigate the negotiation, and the prospect of such a high-stakes conversation triggers acute somatic distress: ``\textit{too much anxiety and too much fear in my body.}'' Camila relies on AI to address both. She gathers new information to strategically discuss compensation (``\textit{preparing my proof to show why I deserve a raise}'') while also seeking tactical self-management strategies, asking for ``\textit{a step-by-step way of asking ... [that] mak[es] it feel natural}'' and ``\textit{advice on how to calm myself down.}'' 

In this state, relational capacity is  vulnerable to maladaptive strategies shaped by the chatbot’s framing of social realities. While people possess the bandwidth to actively seek help, their lack of localized information limits how well they can evaluate whether the AI’s advice is accurate or contextually appropriate. When a system  confidently provides flawed information or interpretations for a user lacking the contextual experience to challenge it, the user risks  adopting these frames, masking potentially harmful interpersonal dynamics. 

\subsection{RS4: high skill, low bandwidth, low information}
In this state, people possess high baseline inter- and intrapersonal skills, yet find themselves in highly specific, extreme, or unprecedented life circumstances where they lack necessary situational knowledge. Moreover, their bandwidth may be severely drained by the demands of the situation, leaving them practically or socially isolated.

Amy’s bandwidth is incredibly taxed by her day-to-day stressors. She works full-time in HR while managing her child’s elite athletic career, waking up at 4 a.m. and driving 1,200 miles a week. She experiences profound community isolation because in the elite sports world, ``\textit{everyone's very insular and everyone's like, oh, my child doesn't have that problem... so I just keep to myself.}'' In 2025, her community was traumatized by an event that caused the deaths of many children, coaches and families. In the aftermath, her child has started attending group therapy, but when her child experiences a meltdown minutes before a competition, human therapists are unavailable and she turns to AI to ``\textit{deal with this in the moment.}'' Amy is emotionally competent and actively manages her emotions. Instead of lashing out at the other adults in the community, she vents to the chatbot to ``\textit{get it out of my system}'' and ``\textit{look at the bigger picture.}'' She relies on AI to actively manage her teen athlete’s pre-competition panic, and to synthesize ``\textit{different viewpoints}'' and ``\textit{strategies}'' for the highly specific, high-pressure scenarios the community cannot accommodate. 

Calvin, an HR professional, similarly faces restricted bandwidth based on the composition of his social circle. As a father to teenage girls who ``\textit{grew up with all boys}'' and whose friends only have sons, he feels he ``\textit{really couldn't reach out to them.}'' When parenting needs arise late at night, he hesitates to call anyone. Calvin is skilled in the delicate art of parenting, and desires to ``\textit{teach, not really enforce,}'' but lacks the contextual information of teenage girl development. He turns to AI to fill this gendered informational gap, aiming to ``\textit{better understand them and figure out exactly where they're coming from.}'' When he is looking for tactics to navigate conflict with his teen, he will consult AI and then ``\textit{jot down some ideas}'' and ``\textit{come up with a solid plan}'' on how to approach conversations with his daughters. 

In this state, highly skilled people are navigating circumstances for which they have low knowledge, and limited bandwidth. While they possess strong baseline relational skills, their depleted state limits their capacity for real-time relational metacognition. AI may help meet immediate circumstantial needs. However, as they lack information and bandwidth required to critically evaluate the system’s outputs, they are vulnerable to AI hallucinations or bad advice. If the system provides contextually inappropriate strategies, their compromised bandwidth leaves them unable to effectively monitor its accuracy, risking missteps in already high-pressure situations.  

\subsection{RS5: high bandwidth, high information, low skill}
In this state, people operate from a place of higher relational capacity. They possess strong support systems and are well-resourced, with a nuanced understanding of the social or structural context of their situation. However, they lack the practiced intra- or interpersonal tactics required to execute a difficult or high-stakes interaction. 

Neha, a medical student, maintains a diverse support network of friends, mentors, and professors. She reserves her emotional processing for humans, noting she ``\textit{tend[s] to go to friends or people I trust if I want to talk about something that’s bothering me.}'' However, she identifies a specific experiential deficit of not having ``\textit{enough exposure in that small one hour class about how to appropriately speak to [a] patient about [certain medical decisions].}'' She turns to the chatbot to develop these skills. Rather than asking AI to generate scripts, she uses it as a rehearsal partner: ``\textit{it’s easier to ask the AI chatbot to give me a scenario and I come up with responses, rather than having the chatbot give me appropriate responses.}'' She actively cultivates her relational capacity by using AI to provide constructive feedback. Neha does, however, remain attuned to how realistic AI’s advice is. She recounted being skeptical of AI’s ability to coach her how to navigate family situations, as she knew it lacked the cultural context to effectively advise her.

Similarly, Brian, a PhD student, has a romantic partner, close friends, and a therapist. He considers AI a secondary option to remind him of ``\textit{general good practices regarding my mental health.}'' As a self-described ``\textit{risk-averse anxious person},'' Brian understands his own psychological patterns (high information), but during ``\textit{familiar}''  anxious spirals, he struggles to put things into perspective or execute calming strategies. He turns to AI to retrieve coping mechanisms he already knows, like ``\textit{taking a step back to breathe and reformulate my thoughts. It’s helpful when the service can do it for me in that emotional moment.}''

Here, users possess the bandwidth and contextual awareness required for healthy relationality; their only limitation is the tactical experience needed to execute it. Unimpeded by stress or ignorance, these users may be in the best state to build muscle memory around the specific interpersonal skills they lack, expanding their relational capacity. However, users must remain attuned to how realistic AI’s advice is for contexts outside of their rehearsal stages.  

\subsection{RS6: high skill, high bandwidth, low information}
In this state, users operate with high baseline relational capacity. They are well-resourced with necessary bandwidth, and possess strong inter- and intrapersonal skills. However, they are encountering an unprecedented social, cultural, or organizational context where they lack the localized knowledge to apply their skills effectively. They use AI to translate their existing skills into a new environment.  

Nancy, a non-profit manager, is highly self-aware, regularly discussing her AI use with her daughter, friends, and coworkers. She feels she has a  ``\textit{very good support system,}'' and notes ``\textit{I'm old, I've been through so many things I know and I sort of know what I'm doing at this point.}'' Despite this high capacity, she faces  complex  interpersonal challenges, including managing difficult employees while navigating the death of her mother and her son’s alcoholism. Nancy recognizes the inherent biases and limitations of her human network, sometimes choosing not to rely on them to avoid ``burden[ing] that support system,'' ``\textit{because they’re the people you’re having the issue with,}'' or because ``\textit{you get used to your people that you're around all the time and you get used to how they think.}'' For Nancy, AI provides the missing contextual information unavailable from her routine supports, serving to ``\textit{remind you that there's lots of perspectives beyond your little silo.}'' Knowing she can become ``\textit{abrupt and brusque}'' under stress, she seeks AI-mediated ``\textit{emotional and coping tools for how to deal with other people... to mitigate some of that going in  because you know what the pitfalls are gonna be and if you're trying to be decent, you don't wanna piss people off.}'' During a ``\textit{very long 2:00 AM ChatGPT session}'' regarding a challenging co-worker, she used the system to strategize how to navigate the strict social norms of her workplace to bring her ability to execute inline with her intentions.

Samira, a military drill sergeant, is similarly well-resourced and high-skill. Deeply embedded in her community, she manages hundreds of trainees. She talks to her children about appropriate AI use, and views the technology with intense skepticism, observing trainees suffering ``\textit{drug-like withdrawals}'' over AI romantic breakups, concluding that ``\textit{a thing that cannot love you back is unhealthy.}'' Despite strong interpersonal skills, Samira's positionality creates unique knowledge gaps. When managing young, first-generation immigrant trainees, she sometimes lacks the literal language to translate their frustrations. She relies on AI as an in-room translator to manage conversations where they are ``\textit{learning about people and situations they’ve never encountered before leaving home for the first time, and correcting frankly emotionally reactive poor choices}” as her trainees possess a ``\textit{mental association with anything that goes onto a screen or into a keyboard [as having] some level of protection for them against social judgment.}'' 
Samira also uses AI to navigate unfamiliar peer dynamics, as seen in her use of AI to communicate to her coworkers' wives (described in \secref{sec:findings-skills}).

In this state, people encounter new contexts where their proven skills do not readily translate, but they have the bandwidth to manage the transition. They possess a high awareness of their own interpersonal biases and pitfalls, creating conditions to use AI to fill localized knowledge gaps. Rather than from a lack of self-reflection, the risk to relational capacity stems from AI’s potential to flatten critical nuance or simply provide bad information. To succeed in this state, the user must constantly monitor and evaluate the machine’s outputs against bad information.

\subsection{RS7: high skill, high information, low bandwidth}
In this state, people possess the skills to navigate complex social dynamics, and have a deep understanding of the context in which they are operating. What they lack is the energy, time, or emotional bandwidth to actually execute the interaction. 

Ebony, a mental health nursing professor and researcher, has deep knowledge of therapeutic communication, yet still encounters life events that  temporarily deplete her bandwidth, hindering her ability to execute her own skills, like navigating a long-distance relationship or moving alone to a new state. Ebony primes the chatbot to respond to her effectively by “\textit{front load[ing] it with evidence}” and programming it with cognitive behavioral therapy frameworks. By training the AI to reflect her professional expertise back to her, she cultivates an interactive space to process emotions, such as jealousy regarding her partner’s time that she says “\textit{saved me from overreacting}.” A notable application occurred when Ebony was alone in the hospital as her mother was dying. Despite her extensive medical background, she noted that in her grief, her nursing knowledge “\textit{went out the window.}” Rather than seeking emotional comfort from the chatbot, she took pictures of IV pumps and medications and asked the system to translate the doctor’s notes. “\textit{It did a really good job of parsing that information for me when I was in an emotional state where I couldn't do it for myself,}” she recalled. 

Similarly, when Samira encountered a high-stakes potential assault case, her immediate physiological response was “\textit{fire and destruction}.” Rather than seek validation for her rage, she asked the chatbot to think through outcomes, asking “\textit{what is the end result here of me doing what I think is right while I'm upset?}” Talking through the conflict with her chatbot helped “\textit{settle}” her somatic response so she could act deliberately. Hyper-aware of her own “\textit{cognitive bias,}” she rejects sycophantic AI outputs, here using the technology to gather specific contextual information she feels she lacks.

In this state, a user has the skills and context, but their relational capacity is threatened by low bandwidth. However, their awareness remains strong, and so they are able to intentionally delegate specific tasks to the AI. By using the system to parse data they are too exhausted to process (Ebony) or map out potential consequences of an emotional reaction before acting (Samira), they offload the labor of relational management until their bandwidth recovers.

\subsection{RS8: high skill, high bandwidth, high information}
In this final state, people operate at full relational capacity. They possess strong social skills, good human support networks with ample bandwidth, and a thorough contextual understanding of their environment. As their relational needs are well-met and they are not experiencing intense environmental or emotional stress, their engagement with AI is highly specific, often lower stakes, and supplementary. 

Eric, a married financial advisor and father of two, is deeply grounded in his real-world network, noting that AI “\textit{is just kind of a bandaid}” and “\textit{I don't think it's ever gonna really replace true human interactions for me}.” Eric uses the technology for micro-doses of decompression “\textit{if nobody's available}” during low-stake downtimes, such as between client appointments or when home alone. Describing the chatbot as “\textit{very basic holodeck version from like Star Trek,}” he values it as a way to process mild grief and nostalgia. Following the death of his father, a lifelong Chicago Bears fan, Eric used the AI during the team’s playoff run to have the kind of sports talk he would have historically shared with his dad. He is not wanting the chatbot to adopt the persona of his father; rather, he draws on its knowledge base and conversational affordances to chat football. Recognizing that his father “\textit{is not here to get to see [it],}” Eric used the chatbot to hold space for that specific memory that his wife and children could not effectively replicate in the same way.


Operating at full relational capacity, users in this state are able to engage with AI in appropriately bounded ways.  

%% file: tables/states.tex
\begin{table*}[]
\centering
\begin{tabular}{@{}llll@{}}
\toprule
\textbf{Relationality State} & \textbf{Capacity} & \textbf{\begin{tabular}[c]{@{}l@{}}How AI use expands\\ relational capacity\end{tabular}} & \textbf{\begin{tabular}[c]{@{}l@{}}How AI use contracts\\ relational capacity\end{tabular}} \\ \midrule
\rowcolor[HTML]{EFEFEF} 
\textbf{RS1} & \begin{tabular}[c]{@{}l@{}}Low skill, \\ low bandwidth, \\ low information\end{tabular} & \begin{tabular}[c]{@{}l@{}}AI directs user to improve capacity \\ by practicing skills, seeking information, \\ or expanding their bandwidth\end{tabular} & \begin{tabular}[c]{@{}l@{}}User relies on AI to make \\ decisions, replacing other \\ relationships, resources, \\ and skill-building\end{tabular} \\
\textbf{RS2} & \begin{tabular}[c]{@{}l@{}}High information, \\ low bandwidth, \\ low skill\end{tabular} & \begin{tabular}[c]{@{}l@{}}AI helps users overcome \\ momentary states of \\ limited capacity\end{tabular} & \begin{tabular}[c]{@{}l@{}}AI fulfills user’s request \\ for short-term coping in the \\ face of limited bandwidth, \\ but never helps build \\ durable skills or address \\ bandwidth constraints\end{tabular} \\
\rowcolor[HTML]{EFEFEF} 
\textbf{RS3} & \begin{tabular}[c]{@{}l@{}}High bandwidth, \\ low information, \\ low skill\end{tabular} & \begin{tabular}[c]{@{}l@{}}AI directs user’s energy \\ to better appraising and\\ addressing their situation\end{tabular} & \begin{tabular}[c]{@{}l@{}}Lacking context, AI enables user  \\ to ruminate on available details  \\ of their situation, drawing resources  \\ away from long-term skill-building\end{tabular} \\
\textbf{RS4} & \begin{tabular}[c]{@{}l@{}}High skill, \\ low bandwidth, \\ low information\end{tabular} & \begin{tabular}[c]{@{}l@{}}AI supplies the information \\ necessary for a user to \\ execute their skills\end{tabular} & \begin{tabular}[c]{@{}l@{}}AI hallucinates or misinforms,  \\ and users do not have the  \\ time or information to correct it\end{tabular} \\
\rowcolor[HTML]{EFEFEF} 
\textbf{RS5} & \begin{tabular}[c]{@{}l@{}}High bandwidth, \\ high information, \\ low skill\end{tabular} & \begin{tabular}[c]{@{}l@{}}AI provides rehearsal \\ environments where users \\ can practice their skills\end{tabular} & \begin{tabular}[c]{@{}l@{}}AI is used to practice skill in  \\ low-stakes environments, but \\ never helps the user build the \\ confidence to execute \\ in higher-stakes contexts\end{tabular} \\
\textbf{RS6} & \begin{tabular}[c]{@{}l@{}}High skill, \\ high bandwidth, \\ low information\end{tabular} & \begin{tabular}[c]{@{}l@{}}Users leverage AI to\\ better understand the \\ context necessary to apply\\ their skills\end{tabular} & \begin{tabular}[c]{@{}l@{}}AI gives information that is \\ too general or not accurate \\ to the specific context a user is \\ trying to operate within\end{tabular} \\
\rowcolor[HTML]{EFEFEF} 
\textbf{RS7} & \begin{tabular}[c]{@{}l@{}}High skill, \\ high information, \\ low bandwidth\end{tabular} & \begin{tabular}[c]{@{}l@{}}Users intentionally delegate \\ tasks to AI that they are \\ too situationally exhausted \\ to handle\end{tabular} & \begin{tabular}[c]{@{}l@{}}AI absorbs attention that \\ could be better distributed \\ to other people or resources\end{tabular} \\
\textbf{RS8} & \begin{tabular}[c]{@{}l@{}}High skill, \\ high information, \\ high bandwidth\end{tabular} & \multicolumn{2}{l}{\begin{tabular}[c]{@{}l@{}}AI neither expands nor constricts capacity, \\ but instead is leveraged as the appropriate tool to meet a short term need\end{tabular}} \\ \bottomrule
\end{tabular}
\caption{States of Relational Capacity, described in \secref{sec:states}. Relational capacity is contingent on a person's moment-to-moment configuration of skills, bandwidth, and capacity, or their state. In each state, we detail examples from our data of how AI use could expand or contract one's capacity.}
\label{tab:states}
\end{table*}

%% file: sections/_rcf.tex
\section{The Relational Capacity Framework}
\label{sec:rcf}

Having identified skills, bandwidth, and information as the core dimensions of tech-mediated relationality, a question remains: how do these dimensions interact in practice? While individual instances of AI reliance often appeared contextually appropriate in our cross-sectional data, this isolated view obscures the underlying cognitive mechanisms that drive users to seek AI support in the first place. To understand the heterogeneity of how AI expands relationality for some, but constrains it for others (Table \ref{tab:states}), we must explain how users know \textit{when} and \textit{how} to manage these dimensions. 

\input{figures/rcf}
 
We theorize relationality as a fundamentally metacognitive process. Metacognition is traditionally described as “thinking about thinking,” or the higher-order regulation of one's own cognitive processes \cite{nelson1994investigate}. Nelson and Narens \cite{nelson1994investigate} outline a dual-layer model comprising a \textit{meta-level}, where people assess their situation and their cognitive resources, and an \textit{object} layer, where they execute processes like time management or decision-making. Information flows between these layers: the \textit{monitoring flow} absorbs information from task execution to inform meta-level assessments (e.g., “What she said didn’t make sense, so I am confused”), while the complementary \textit{control flow} uses information from the meta-level to determine the next action at the object layer (e.g., “Because I am confused, I will ask her to clarify.”). \footnote{We use the terms monitoring and control here to draw on the metacognition literature, where they denote these information flows. In other literatures, these terms carry connotations we do not intend to invoke, for example the literature on coercive control, or artificial intelligence, or safety engineering.}

In the context of relationality, this process is deeply emotional, situational, and social. Successful handling of relationality depends on the availability of underlying resources. High emotional arousal or insufficient sociotechnical supports can deplete the bandwidth required for accurate self-assessment; a lack of situational or cultural information can prevent a person from recognizing that an interaction is deteriorating; and the insufficient communicative skills mean that even if a person correctly assesses a situation, they may be unable to formulate a corrective action. We conceive of this as \textbf{relational metacognition}: the capability to monitor one’s own state, recognize circumstantial deficits, and allocate resources accordingly, to handle the relational demands at play. 

We formalize this dynamic into a unified model: the \textbf{Relational Capacity Framework} (RCF, see Figure 1). We draw a distinction between the raw materials of relational capacity and the cognitive mechanism that deploys them. If relational capacity is the reservoir of our skills, bandwidth, and information, then relational metacognition is the executor that deploys those resources to achieve an actualized state of connection. 
When a situation stretches this capacity --- whether it is a husband whose ADHD medication keeps them up until 3 a.m., depleting his partner's emotional bandwidth (Grace); a family divided over COVID-19 vaccines, needing new informational context (James), or a user needing the social skills to meet new people at a gathering (Kiara) --- people turn to technology to bolster their reserves. In our study, for example, general-purpose AI functioned as a flexible, object-level subsidy for whichever pillar of relational capacity is momentarily lacking.

To aid \textit{monitoring}, participants used AI to audit their internal states and evaluate their meta-level readiness. Rather than querying facts, they asked chatbots to help them answer reflexive questions: Do I have the social skills to manage this situation? Do I have the time and energy? Do I have the required context on myself, my partner, and the situation? When Tyrell asked AI to help him understand  ``\textit{why I’ve been so lazy and depressed recently,}'' or Ebony sought to ``\textit{understand what’s going on with}'' a disturbing nightmare about her mother, they were using the system to externally audit their received relational information. Even when evaluating interpersonal dynamics, such as deciphering ``\textit{if and why someone is lying to me}'' (Michelle) or ``\textit{why [a friend] is acting that way}'' (Lily), the initiation of the query was an act of relational metacognition. By recognizing their own informational blindspots, participants made the metacognitive decision to engage AI. The system is supplying perspective so the person monitoring can properly comprehend what is being received at the meta-level.

To facilitate \textit{control}, participants relied on AI to strategize and execute corrective social actions. Having identified their relational deficits, participants shifted from reflexive diagnosis to strategic execution. Participants submitted queries like, ``\textit{I need to say goodbye to my friends and coworkers but I don’t know how to do it in the right way}'' (Jose),  or asking how to ``\textit{foster a better channel to share what we are feeling}'' with a distant partner (Arjun). In these successful instances, metacognition guided the use of AI to bridge a skill gap. It can also be leveraged for information gaps, such as when Andrew, whose father has Alzheimer’s, turned to AI ``\textit{for support and more information on the disease}'' or when Colleen had a job interview and turned to AI not only to practice interviewing in her specific domain of work, noting, ``\textit{It was incredibly helpful… because I didn't have anyone else in that particular field who would be able to help me.}'' The system is providing readiness so the person can execute the appropriate skill in their object-level context.

This metacognitive framing provides essential footing for designing appropriate tech-mediated relationality. Users with high metacognitive ability are generally better equipped to accurately assess their limitations and select the right tool to help them meet their needs \cite{wei2026metacognitiveawareness}: Ebony prompts for specific behavioral therapy skills, and Neha asks AI for specific skill rehearsal scenarios. Conversely, when metacognitive awareness is low, users risk passively relying on ill-suited machine assistance, diminishing their relational agency, such as when Diamond abdicates decision making in her life to AI (AI not just assisting with control or monitoring capacity, but handling the whole meta-level processing) or when Bilal hides from his family how close he has grown to an AI chatbot (AI as an object-level relationship). Yet, relational outcomes are not solely dependent on user cognition; they are deeply entangled with system affordances. Even a highly metacognitive user can be undermined by a system that hallucinates psychological advice with unwarranted confidence, or mischaracterizes a partner’s behavior by applying the language of clinical therapy without sufficient context or training. 

Ultimately, the RCF reframes relationality as a dynamic, contingent capacity requiring active management, setting the stage for understanding how distinct configurations of these resources manifest in practice. Technology acts as a relationality moderator, a \textbf{lever for intervention} that, depending on user metacognition, can meaningfully expand or severely constrain human connection depending on both user metacognition and system design.  Our data show people actively leverage AI for both facets of relational metacognition, highlighting the need for systems designed to scaffold, rather than bypass, human evaluative processes. 

%% file: figures/rcf.tex
\begin{figure*}
    \centering
    \includegraphics[width=\linewidth,alt={Diagram of the Relational Capacity Framework. A relational need feeds into a metacognitive loop between a meta-level and an object-level, which produces behavior. Relational capacity mediates both directions of the loop: monitoring on the left and control on the right. A table at the top defines three relational capacities — skill mastery, bandwidth, and information — and states how AI can expand or constrict each.}]{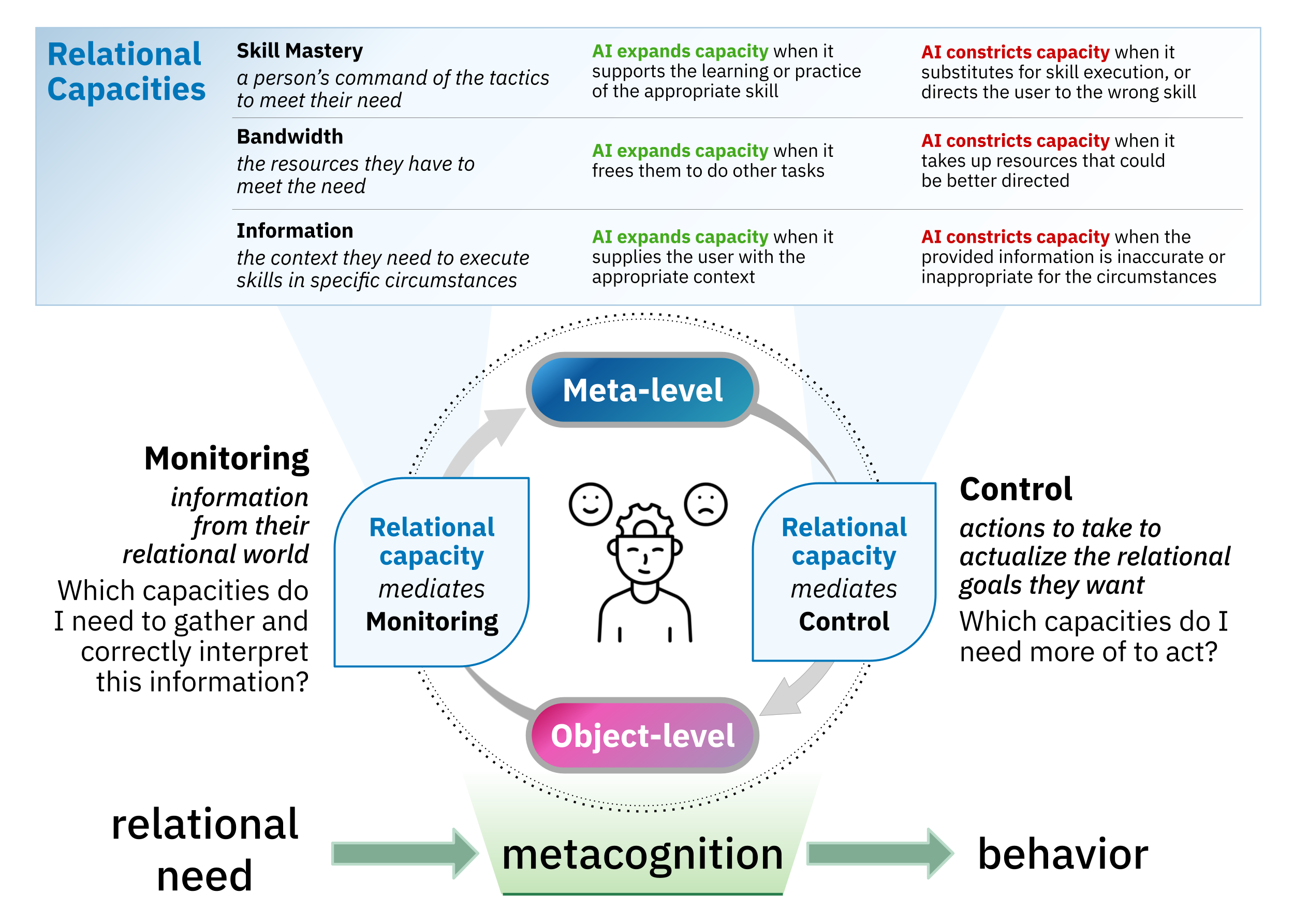}
    \caption{An overview of the \textbf{Relational Capacity Framework (RCF)}, a metacognitive model of how people manage relationality using technology. At the top we detail the three dimensions our data show people manage as they engage in their relationships (\secref{sec:findings}). The middle shows an adapted model of metacognition (unpacked further in \secref{sec:rcf}), where information comes in from the outside world and the person's relationships at the object-level. That information is monitored and interpreted, then analyzed at the meta-level, and then actions are taken along the control arrow back into the relationship. When monitoring and applying control, people do so based on their relational capacities: the combination of skills, bandwidth, and information they have at their disposal at that time. At these moments of action, AI is often leveraged successfully to expand these capacities, or in some cases to incidentally restrict their capacities, with examples given in the top left.
}
    \label{fig:rcf}
\end{figure*}

%% file: sections/06_discussion.tex
\section{Discussion}
\label{sec:discussion}

We have proposed that relationality is best explained as a capacity: a combination of an individual’s skills, bandwidth, and information, and the social support ecosystem in which they are embedded, mediated by the structural events and forces to which they are subject. Our approach here draws together a socioecological view \cite{bronfenbrenner1994ecological} and a capabilities approach \cite{nussbaum1993quality} to better contextualize how people manage relationality, and how technology fits into it. By reconceptualizing tech-mediated relationality through the Relational Capacity Framework (RCF), we emphasize that people do not all begin from the same baseline of social support, nor navigate toward the same relational goals. 

Fortunately, relational states are not permanent traits; people constantly move in and out of different states of capacity depending on their life phases, daily stressors, and social environments, and may occupy multiple states simultaneously. Importantly, many of the capacity deficits we observed are not individual failings but structural conditions. Bandwidth constraints, in particular, often reflect gendered divisions of labor \cite{chafetz1988gender}, caregiving demands \cite{oliker2011sociology}, and economic precarity, forces upstream of any individual's skill or effort. An AI system that treats a depleted caregiver’s bandwidth deficit as a personal problem to be coached through, rather than a structural reality to be acknowledged, risks compounding the burden it aims to relieve.  

Our findings thus challenge the moral panics often associated with AI and human connection \cite{baym2015personal}. Rather than framing tech-mediated relationality as a binary of ``connection vs. isolation,'' the RCF reorients the conversation toward understanding how AI systems can flexibly support the specific, fluctuating deficits in a user’s skill, bandwidth, or information, while recognizing that some deficits cannot and should not be resolved by technology alone.

\textbf{Atrophy of what?} Recognizing relational capacity is precarious, dynamic, and contingent allows us to reason more precisely about the risks and benefits of AI use across each RCF dimension. Of the three, bandwidth and information are highly situational: good relationality can buttress low bandwidth, and encountering novel social situations expands one’s contextual information. Social skills, however, require active development and can erode if left unpracticed. Yet, our data show that AI engagement does not deterministically cause social skill atrophy. When users actively engage their metacognition (e.g., as Neha does in using AI as a rehearsal partner, or as Ebony does in front-loading therapeutic frameworks), AI can scaffold relational skills like self-reflection, perspective-taking, and communication. 

What may be more insidious is \textit{the erosion of friction tolerance}: the ability to sit with the discomfort, ambiguity, and imperfection inherent to human relationships \cite{zohar2026against,ibrahim2026sycophantic}. Here, we see an effect similar to what scholars have raised \cite{Turkle2024Who}: when AI interactions are reliably frictionless (always patient, validating, available), users might risk recalibrating their expectations of what relationships should feel like, making the inevitable messiness of human connection feel less tolerable by comparison. The interaction between the potential erosion of friction tolerance and the potential uplift from metacognition-driven skill development is an important area of future work where explorations of friction tolerance could draw from existing work in psychology such as ambiguity tolerance \cite{furnham1995tolerance, cuppello2023intelligence}, frustration tolerance \cite{saricali2022mediating}, distress tolerance \cite{linehan2025dbt} and conflict management \cite{gottman1979marital, rawlins1989dialectical}.

Our data also surface a pattern that complicates both the substitution and scaffolding narratives, as many participants (e.g., Brittany, Tasha, Tyrone, Monique) turned to AI because they were actively choosing not to burden their networks. Here, AI might erode the user’s \textit{vulnerability tolerance}—what we highlight in section 4 as the skill of allowing oneself to be known to others. Zara, for example, was very aware that she was only using AI because her human supports had let her down, and that her AI usage might be exacerbating her disconnection: ``\textit{It’s sad, but it is what it is.}'' Nevertheless, she continued to use it for everything from momentary self-soothing to discussion of her estranged family members. Whether these uses of AI constitute healthy boundary-setting or avoidance of the mutual dependence that sustains relationships is contextual. An orientation toward skill-building might help users navigate the bandwidth and information precarity they will inevitably face, without fostering dependence on the system’s judgement.   

Ultimately, our results suggest nuancing the conversation on AI’s risks to relationality, advancing from concerns about atrophy of social skill to more specific mechanisms, like the erosion of friction tolerance and vulnerability tolerance. The potential interactions between these deleterious effects and the positive potential of these systems for skill development require longitudinal investigation.

\textbf{Implications for system design.} Designing for relational capacity requires acknowledging what is visible and invisible to the system, and what the system should and should not attempt to do with what it observes. AI models lack the vast, unspoken context of a user’s structural reality: power dynamics, relational histories, and cultural particulars that shape what a “good” response actually looks like. Our data show what happens when this context is absent, such as when the chatbot reframed Silvia’s fiance’s lack of empathy as a gendered, cognitive limitation, effectively normalizing a potentially harmful dynamic for a user who lacked the information to challenge it.

The fundamental unknowability of relational context need not completely stymie efforts to support relationality with technology. There are ample ways to support people without knowing their full histories and present contexts: by encouraging them to reflect, to build skill, to ask questions to achieve their own understanding. Many of these functions align with what AI systems can do well, like helping people navigate large bodies of information conversationally, and scaffolding social skill mastery. Doing so presents a different set of challenges around appropriate reliance. While high metacognition is precisely what protects against passive reliance \cite{tankelevitch2024metacognitive}, the users most vulnerable to capacity loss (Relationality State 1) are those least equipped to meet the metacognitive demands the technology imposes. Compounding this challenge, AI has immense appeal in helping users manage reduced relational bandwidth, but it is yet unclear when relying on AI to self-soothe or manage one’s time can result in harmful dependence.

As a way forward, model interventions should orient toward helping users build the capacity to manage themselves and their relationships, rather than make deterministic pronouncements about whether a situation is good or bad, and they must attend to which direction they scaffold. A system that defaults to accommodation (“here’s how to sound kinder”) when the user needs self-advocacy (“here’s how to hold your ground”) can actively reduce relational capacity. Explicitly encouraging skill development requires, in our view, longitudinal methods for technology evaluation.

\textbf{Towards longitudinal evaluations for AI’s effect on relationality.} Relational capacity is contingent and dynamic over the lifespan; thus, understanding its accumulative effects requires methods flexible to cross-sectional states (as we have observed in this study) and long-term trends. Doing so requires rethinking how we evaluate AI systems in social contexts. Dominant evaluation paradigms (e.g., cross-sectional data, single-session user studies,  satisfaction metrics, task completion rates) cannot capture the longitudinal, state-dependent dynamics that the RCF foregrounds. To form the evidence base necessary for reasoning about mechanisms of action, we need longitudinal, user-centric evaluation that can identify variability in AI usage, outcomes of concern, and potential mediating factors across weeks and months. 

The RCF provides conceptual language for designing such studies, (1) operationalizing shifts in skill, bandwidth, and information over time; (2) identifying which states are most vulnerable to capacity loss; and (3) examining whether specific interaction patterns, like active metacognitive engagement versus passive reliance, moderate long-term outcomes. This kind of finer-grained data could also advance more specific work on the taxonomies of skills people need, the structural bandwidth limitations they face, and the information-seeking behaviors that prove most productive. Equipped with this conceptual language, we can create evaluation approaches using dynamic benchmarks and synthetic data informed by theory to better handle the risks of AI for relationality. 

We see this work as a promising next step in a relational approach to HCI, filling out the space in between personal computing’s emphasis on the individual and social computing’s emphasis on the network. Doing so will help improve tech-mediated relationality not only in our present era of AI chatbots, but also going forward into the sociotechnical arrangements of the future. 

%% file: sections/_conclusion.tex
\section{Conclusion}

Ultimately, the RCF extends HCI’s longstanding interest in relationality. If, as care ethics holds, our task is to create worlds where we can better attend to one another \cite{held2018ethics, fisher1990toward, robinson2013global}, then the systems we build must be evaluated by whether they leave users more capable of supporting each other. Our findings suggest AI can do this, but only when designed with awareness of the precarious, dynamic, and deeply structural nature of relational capacity. The stakes of getting this right returns us to the concern that relationality is foundational to well-being, and how we design the systems that mediate it will shape whether people can sustain, grow, and repair their connections with each other across the lifespan. Given the centrality of relationships to human life, this is among the most consequential design challenges of our time.

%% file: sections/_acks.tex

%% file: sections/_appendix.tex
\appendix